\documentclass[journal=jacsat,manuscript=article]{achemso}
\usepackage[version=3]{mhchem} 

\author{Adeel Afridi}
\affiliation{ICFO Institut de Ciencies Fotoniques, Mediterranean Technology Park, 08860 Castelldefels (Barcelona), Spain}
\alsoaffiliation{Nanophotonic Systems Laboratory, Department of Mechanical and Process Engineering, ETH Zurich, 8092 Zurich, Switzerland}
\author{Jan Gieseler}
\affiliation{ICFO Institut de Ciencies Fotoniques, Mediterranean Technology Park, 08860 Castelldefels (Barcelona), Spain}
\author{Nadine Meyer}
\affiliation{Nanophotonic Systems Laboratory, Department of Mechanical and Process Engineering, ETH Zurich, 8092 Zurich, Switzerland}
\author{Romain Quidant}
\affiliation{Nanophotonic Systems Laboratory, Department of Mechanical and Process Engineering, ETH Zurich, 8092 Zurich, Switzerland}
\email{rquidant@ethz.ch}
\title{Ultra-thin Tunable Optomechanical Metalens}
\keywords{Optomechanics, Reconfigurable metasurface, Metalens, \LaTeX}
\begin{document}
\textbf{Reconfigurable metasurfaces offer great promises to enhance photonics technology by combining integration with improved functionalities. Recently, reconfigurability in otherwise static metasurfaces has been achieved by modifying the electric permittivity of the meta-atoms themselves or their immediate surrounding. Yet, it remains challenging to achieve significant and fast tunability without increasing bulkiness. Here, we demonstrate an ultra-thin tunable metalens whose focal distance can be changed through optomechanical control with moderate  continuous wave intensities. We achieve fast focal length changes of more than 5\% with response time of the order of 10\,$\mu$s}.\\

\noindent\textbf{Keywords: Optomechanics, Reconfigurable metasurface, Metalens}\\

A metalens is a two dimensional (2D) metasurface \cite{kildishev2013planar,Yu2014,genevet2017recent} that controls the amplitude, polarization and phase of the impinging light using engineered sub-wavelength resonators, also known as meta-atoms (MA), judicially arranged in periodic or quasi-periodic arrays \cite{Yu2011a,Yu2014,Khorasaninejad2017a,She2018}. To date, metalenses are the center of intense research efforts and have shown great potential to replace bulky optical elements, sometimes even with superior performance \cite{Moon2020}. However, the intrinsic passive nature of metalenses limits their use where active operation is required, such as adaptive vision and imaging \cite{lee2020metasurfaces,khorasaninejad2016metalenses,lalanne2017metalenses,Khorasaninejad2017a,tseng2018metalenses,Bosch2021a,Moon2020,She2018}. To address this challenge, various approaches have emerged that rely on changing the optical properties
of the meta-atoms themselves or their surrounding medium \cite{She2018,Ee2016,Kamali2016,Arbabi2018,Bosch2021a,Shen2020,Fan2020a,Badloe2021,Archetti2022a,Afridi2018,Yao2021,Wang2018,Groever2018,Fu2019a,Yu2020a,Shalaginov2021}. In particular, promising advances were reported using electro-optic control \cite{Bosch2021a,Shen2020,Fan2020a,Badloe2021}, temperature-induced effects \cite{Archetti2022a,Afridi2018}, light-induced effects \cite{Yao2021,Wang2018,Groever2018,Fu2019a,Yu2020a}, phase change media \cite{Shalaginov2021} and mechanical actuation \cite{She2018,Ee2016,Kamali2016,Arbabi2018}. Despite this progress, it remains challenging to combine compactness, fast operation speed and low energy consumption. For instance, phase change material-based metasurface provides large tunability and ultrafast switch ON time (up to MHz) but suffer from long switch OFF time \cite{Mikheeva2021}. Similarly, electrically controlled liquid crystal (LC) embedded metalenses operate at low driving voltages ($<$ 10V). However, they require bulky LC chambers and are polarization dependent due to the birefringence of the LC \cite{yang2022active,Bosch2021a}. MEMS-based technologies promise large focal length changes and high switching speed but require high driving voltages \cite{She2018,Arbabi2018}. On the other hand, mechanical stretching requires bulky mechanical arms limiting integration and response time \cite{Kamali2016}.

Optomechanics \cite{gartner2018integrated,moura2018centimeter,burgwal2020comparing,la2022nanomechanical,burgwal2022enhanced,norte2016mechanical}, where optical forces mediate the interaction between light and structural mechanics, offers new possibilities for reconfigurable metasurfaces. When considering suspended meta-atoms patterned in a thin membrane, the resonantly enhanced electromagnetic forces at resonance can exceed elastic forces of the material, thereby inducing mechanical deformation. Already a small deformation alters the phase delay incurred by the impinging light substantially \cite{Zheludev2016a}. First proposed theoretically \cite{Zhang2013}, the concept of optomechanical metasurfaces features giant nonlinearity, optical bistability and asymmetric light transmission. Shortly after, optomechanically-induced modulation of light transmission through a metasurface  has been experimentally demonstrated \cite{Karvounis2015a}. Similarly, gold meta-atoms supported by pairs of free-standing silicon nitride strings exhibit individual optomechanical plasmonic resonances \cite{ou2018optical}. More recently, an experimental study of a suspended silicon carbide (SiC) metasurface supporting multi-mode vibrational resonances was reported \cite{Ajia2021}. These observations suggest optomechanical control as a promising approach toward compact, fast and low power tunable metalenses.

\begin{figure*}[ht]
\centering
\includegraphics[width=0.6\textwidth]{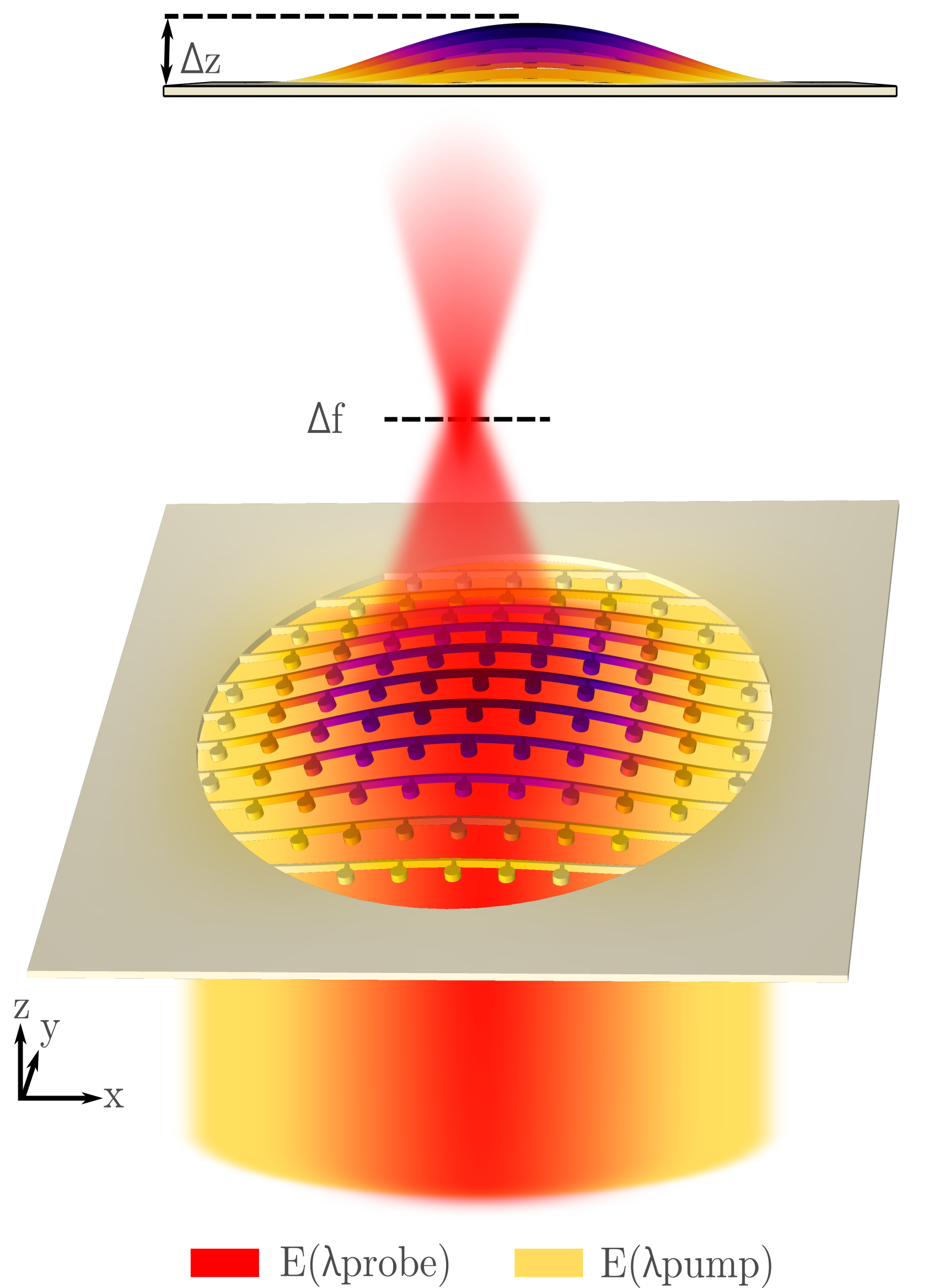}
\caption{Artistic representation of an optomechanically reconfigurable metalens. The metalens is designed to focus light at $\lambda_{probe}$. Upon illumination with a pump beam at $\lambda_{pump}$, the resonantly enhanced optical forces mechanically deforms the metalens by an amount $\Delta z$ resulting in a change of the focal length ($\Delta f$) at $\lambda_{probe}$.}
\label{fig:fig1}
\end{figure*}

In this letter, we design, fabricate and characterize an ultra-thin optomechanically reconfigurable metalens operating in the near-IR regime. Our metalens is formed by an ensemble of suspended Huygens' meta-atoms \cite{decker2015high} carved in a free-standing crystalline silicon membrane, as illustrated in Figure \ref{fig:fig1}. The metalens is designed to focus probe light at $\lambda_{probe}=1.31\,\mu$m while being controlled by a pump beam at $\lambda_{pump} = 1.55\,\mu$m. Optical forces experienced by the meta-atoms translate into a mechanical deformation of the metalens ($\Delta z$) and consequently a change in the focal length $\Delta f$. Beyond demonstrating the tunability of the focal length by the pump light, we perform a full characterization of the lens performance including power dependence, focusing efficiency and time response.

\begin{figure*}[ht]
\centering
\includegraphics[width=0.8\textwidth]{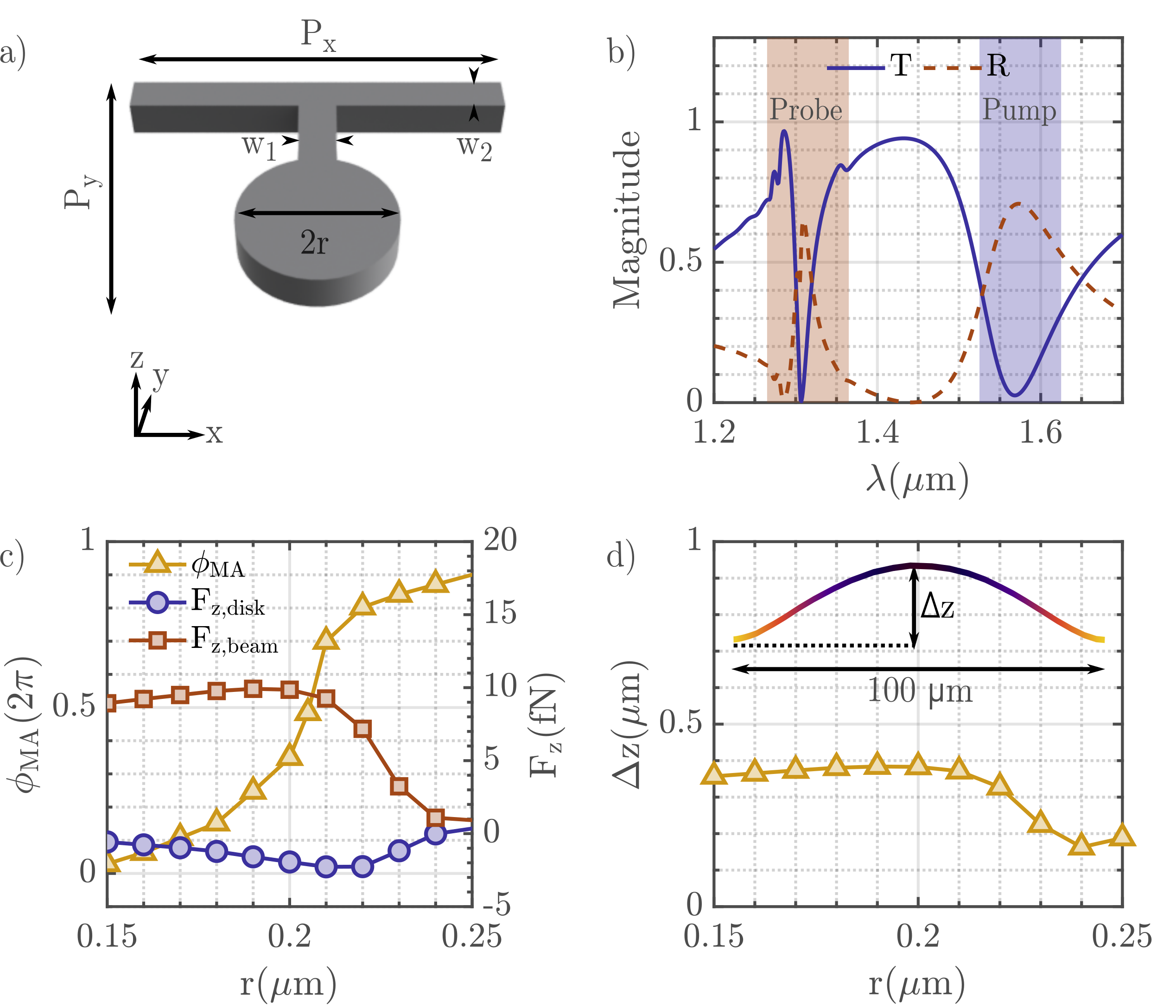}
\caption{Geometry and simulation results of the silicon meta-atom. \textbf{a}) Meta-atom geometrical parameters. \textbf{b}) Simulated transmission \textit{T} (blue solid line), and reflection \textit{R} (red dashed line) of a meta-atom with disk radius $r = 0.2\,\mu$m. The red (blue) shaded region signifies probe (pump) resonance. \textbf{c}) Simulated phase due to the meta-atoms at probe wavelength $\lambda_{pump} = 1.31\,\mu$m and calculated force experienced by the meta-atoms at pump wavelength $\lambda_{probe} = 1.55\,\mu$m and pump intensity $I_{p}= 1$\,$\mu$W/$\mu$m$^2$, as a function of disk radius $r$. \textbf{d}) Numerical simulation of the deformation $\Delta z$ for meta-atoms in a one dimensional array in the presence of pump light with $I_{p} = 110$\,$\mu$W/$\mu$m$^2$.}
\label{fig:fig2}
\end{figure*}

Combining light focusing with optomechanical control requires careful multimode engineering at both pump and probe wavelengths. On the one hand, the set of meta-atoms must cover phase changes from 0 to 2$\pi$ at $\lambda_{probe}$. On the other hand, an additional mode at $\lambda_{pump}$ is required to induce optical forces causing a sufficiently large mechanical deformation. To meet these requirements, we selected a rectangular periodic arrangement of suspended Huygens' meta-atoms with periodicity $P_x$ ($P_y$) along the $x$ ($y$) axis, in a 200\,nm-thick silicon membrane. Each meta-atom consists of a disk with radius $r$, attached to a transversal nanobeam of width $w_2$ via a short neck of width $w_1$ (Figure 2a). Design optimization used a commercial finite element method simulator (COMSOL MULTIPHYSICS). Figure \ref{fig:fig2}b displays the transmission \textit{T} and reflection \textit{R} for a specific meta-atom with \textit{r} = 0.2 $\mu m$, $w_1 = w_2 = 0.095\,\mu $m, $P_x = P_y = 1.275\,\mu$m. Under $x$-polarized light, the disk supports magnetic and electric dipolar resonances around $\lambda_{probe} =  $1.31\,$\mu$m. The interaction between the electric and magnetic dipole resonances, controlled by the disk radius $r$, introduces a phase delay of the incoming light \cite{decker2015high,Arseniy2015}. Additionally, the disk attached to the supporting beam supports an electric mode around $\lambda_{pump}$, that we use for optomechanical control. 

Next, we optimized an entire set of meta-atoms to provide 0-2$\pi$ phase change at $\lambda_{pump}$, by changing the disk radius \textit{r} and beam width $w_2$ simultaneously. The simulated phase imparted ($\phi_{MA}$) by these meta-atoms at $\lambda_{probe}$ are shown in Figure \ref{fig:fig2}c as a function of \textit{r} (yellow solid line with triangular markers), while the respective $w_2$ parameter is given in Table 1 in the supplementary information. The periodicity and the width were kept constant at $P_x = P_y = 1.275\,\mu$m and $w_1 = 0.095\,\mu$m, respectively.

In addition, we used the time averaged Maxwell's stress tensors to calculate the optical forces acting on the meta-atoms, as a function of the disc radius $r$ (see supplementary information for details). Figure 2c shows that the sub-unit (see Figure S3a in supplementary information) of the meta-atoms experience anti-parallel forces along the $z$-direction. The disk experiences a negative force that pulls against the illumination direction. Conversely, the beam is pushed by the incoming light.
This force imbalance across the meta-atom introduces a tilt of the disk with respect to the beam, while the total positive force over a finite periodic array bends the whole array forward. Furthermore, we numerically simulated the deformation $\Delta z$, for a maximum pump intensity of $I_{p} = 110$\,$\mu$W/$\mu$m$^2$ for each meta-atom arranged equidistantly in a 1D array of 100\,$\mu$m \cite{Zhang2013}.  Figure \ref{fig:fig2}d shows the deformation $\Delta z$ as a function of disk radius $r$. We observe a maximum deformation $\Delta z \approx 380\,$nm for the meta-atom with $r = 200$\,nm.

To calculate the required phase profile $\phi_{req}(x,y)$ in the $xy$-plane, that focuses light at focal length $f$, we used the hyperboloidal phase function:
\[\phi_{req}(x,y) = \frac{2\pi}{\lambda_{probe}}(\sqrt{x^2+y^2+f^2}-f)\]
We fixed the focal distance and diameter of the metalens at $f = 300\,\mu$m and $D = 100\,\mu$m, respectively. We discretized the continuous $\phi_{req}(x,y)$ with a suitable meta-atom such that, at a given $x$ and $y$ position, $|\phi_{MA}-\phi_{req}(x,y)|$ is minimized.

\begin{figure*}[ht]
\centering
\includegraphics[width=0.8\textwidth]{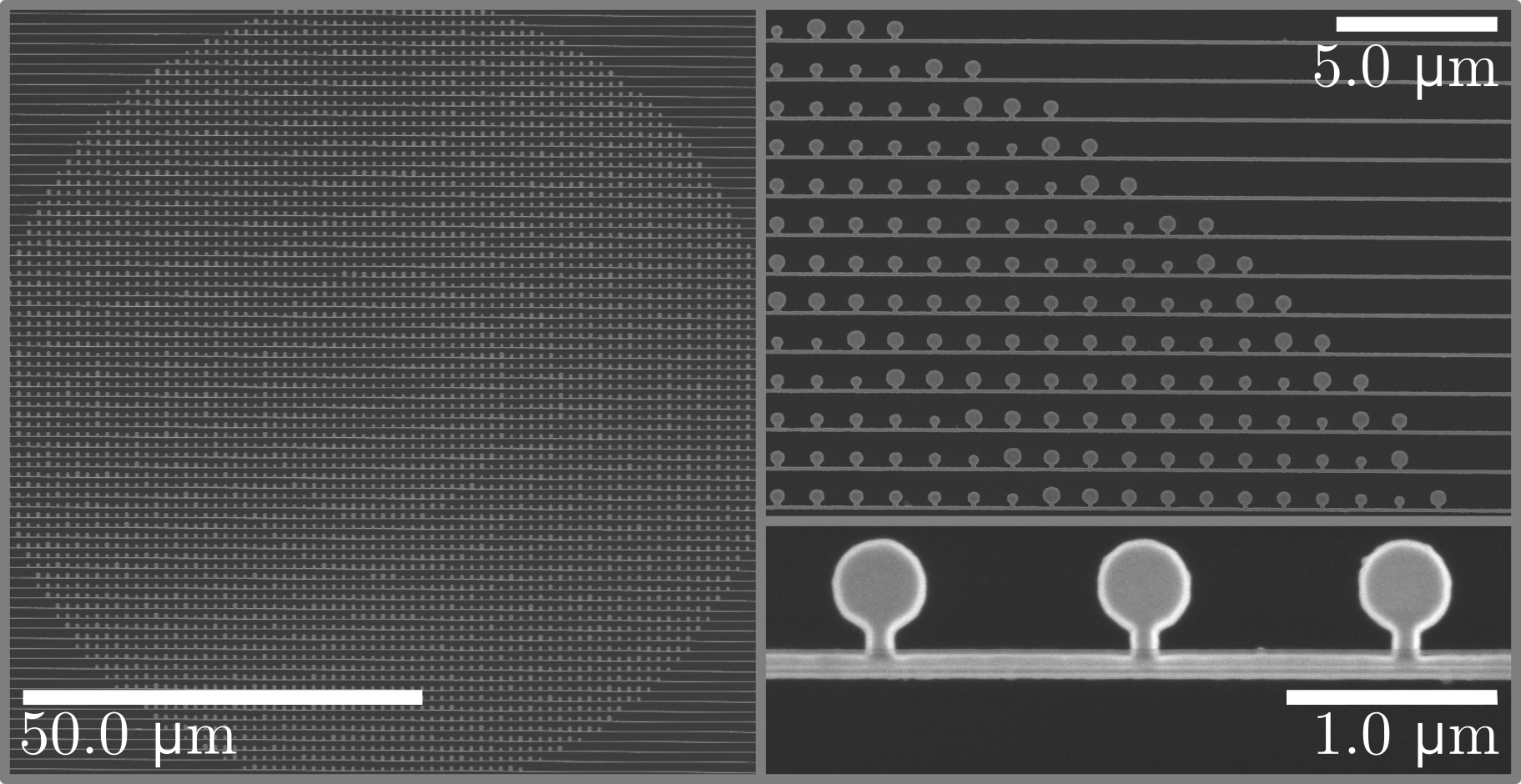}
\caption{SEM micrographs of an optomechanically reconfigurable metalens. Fabricated metalens with diameter $D = 100\,\mu$m and designed focal length $f = 300\,\mu$m carved on a free standing silicon membrane (Norcada Inc.) of thickness 200\,nm.}
\label{fig:fig3}
\end{figure*}

To fabricate our tunable metalens design, we employed top-down electron beam (E-beam) lithography. As material substrate we used a commercially available free standing crystalline (100) silicon membrane from Norcada Inc. We span coated the membrane sample with the AR-P 6200.04 positive photo-resist with a thickness of 230\,nm followed by baking for 1 minute at 150$^o$C. Afterwards, E-beam exposure was carried out followed by one and a half minutes development in AR 600-546 developer at room temperature. We then etched the silicon membrane using HBr chemistry with an inductively coupled plasma (ICP) etcher. Finally, we stripped off the photo-resist with an oxygen plasma etcher. Figure \ref{fig:fig3} shows scanning electron microscopy (SEM) images of the final metalens with a diameter $D = 100\,\mu$m and designed focal length $f = 300\,\mu$m.

\begin{figure*}[ht]
\centering
\includegraphics[width=0.8\textwidth]{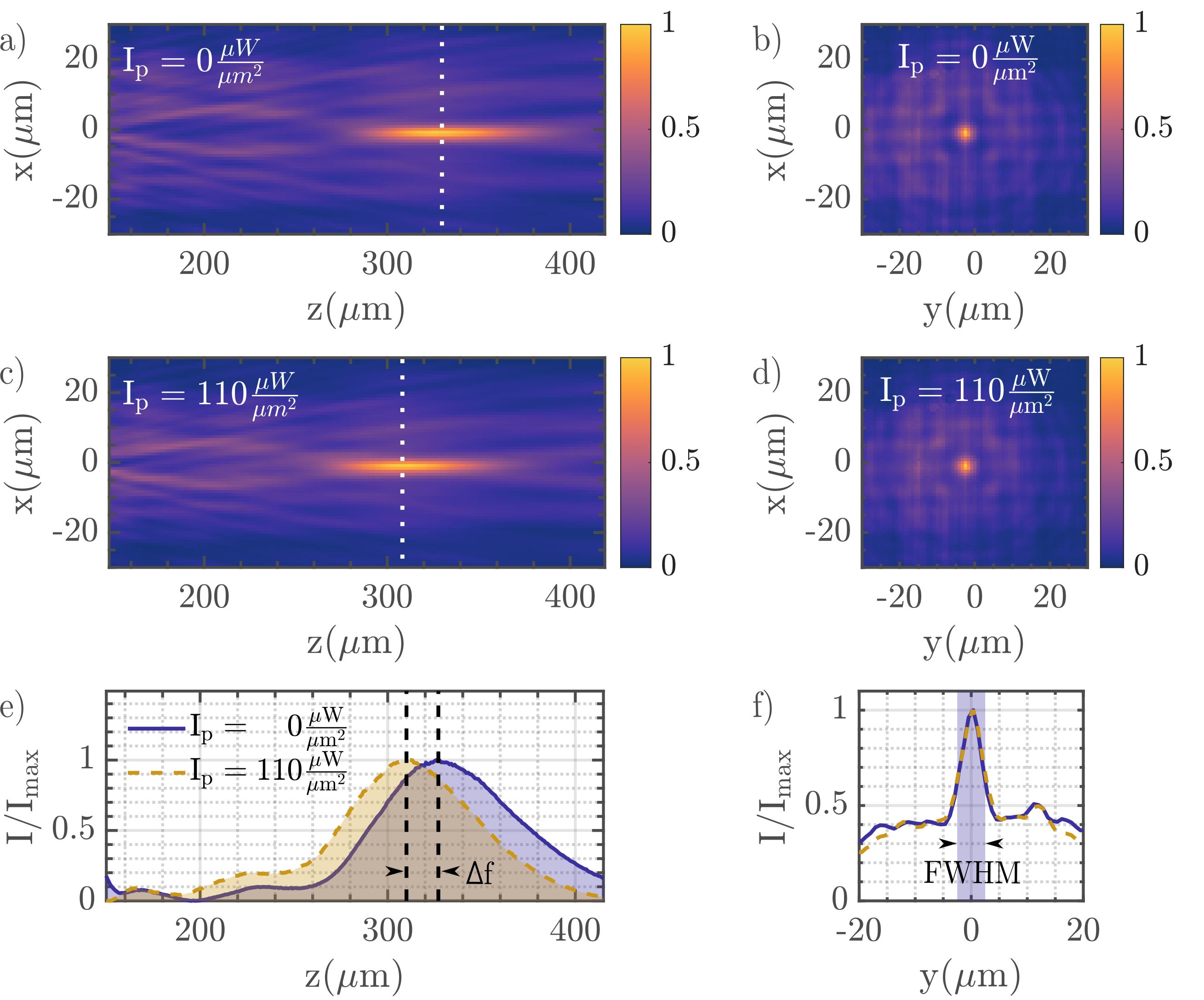}
\caption{Optical characterization of the focal length with pump light ON and OFF. \textbf{a}) and \textbf{c}) 2D intensity maps in the $xz$-plane with the pump intensities $I_{p} = 0\,\mu$W/$\mu$m$^2$ and $I_{p} = 110\,\mu$W/$\mu$m$^2$, respectively. \textbf{b}) and \textbf{d}) 2D intensity profile of the focal spot in its respective focal plane ($xy$) under the pump intensities $I_{p} = 0\,\mu$W/$\mu$m$^2$ and $I_{p} = 110\,\mu$W/$\mu$m$^2$, respectively. \textbf{e}) Axial optical intensity distribution for the two pump intensities. \textbf{f}) 1D cut of the focal spot across the $y$-axis for the two pump intensities (blue solid line for $I_{p} = 0\,\mu$W/$\mu$m$^2$ and yellow dashed line for $I_{p} = 110\,\mu$W/$\mu$m$^2$. The FWHM is highlighted by the blue shaded area. All  intensity profiles are normalized to the maximum probe intensity for the given pump intensity.}
\label{fig:fig4}
\end{figure*}

We characterized the fabricated metalens using a home-made two color optical setup (see Figure S1 of the supplementary information). Collimated probe and pump beams are linearly polarized. The pump beam passes through a combination of an electro opto modulator (EOM), a polarizing beam splitter and a half wave plate to control the power and polarization of the pump beam. Afterwards, both beams were recombined on a 50:50 beam splitter. A 20x objective lens and NA = 0.4 focuses the probe and pump beam to a focal spot of 100\,$\mu$m and 60\,$\mu$m onto the metalens. A second identical objective mounted on a piezoelectric stage collects the light after the metalens. A band pass filter with central frequency of 1.31\,$\mu$m blocks the pump while allowing the probe to reach a near infrared (NIR) camera and a photo-diode. We acquired an image stack of 2D intensity maps along the principle optical axis with a step size of $\sim1\,\mu$m by moving the collection objective with the piezo stage. First, we characterized the lens response in the absence of the pump beam. Figure \ref{fig:fig4}a and b show the 2D intensity map without pump  ($I_{p} = 0$\,$\mu$W/$\mu$m$^2$) in the $xz$-plane (longitudinal) and at the focal plane (transversal), respectively. The measured focal length $f$ is 328\,$\mu$m $\pm$ 0.3\,$\mu$m. Subsequently, we switched on the pump beam (intensity $I_{p}$ = 110 $\mu W/\mu m^2$) and observed a reduction of the focal length by $\Delta f = -20\,\mu$m. The 2D intensity maps (longitudinal and transversal) are displayed in Figure \ref{fig:fig4}c and d. For further comparison, we plot the focus profile ($z$-cut) and 1D focal spot profile with and without pump laser in Figure \ref{fig:fig4}e and f, respectively. It is noteworthy, that the focusing quality and the full width half maximum (FWHM) value is well-preserved upon the introduction of the pump laser. Furthermore, the involved pump intensities are compatible with LED arrays, thus eliminating the need for a bulky laser source.

Although our metalens was optimized for $x$-polarized pump light, we also characterized the metalens under $y$-polarized pump light and find a focal change $\Delta f = -6\,\mu$m for the pump intensity $I_p = 110\,\mu$W/$\mu$m$^2$ (see Figure S2 in supplementary information). This is consistent with numerical simulations shown in Figure S3a, which reveal that under $y$-polarization the effect of the pump on the beam becomes negligible and, in contrast to the $x$-polarized case, the force on the disk is along the propagation direction. 

\begin{figure*}[ht]
\centering
\includegraphics[width=0.8\textwidth]{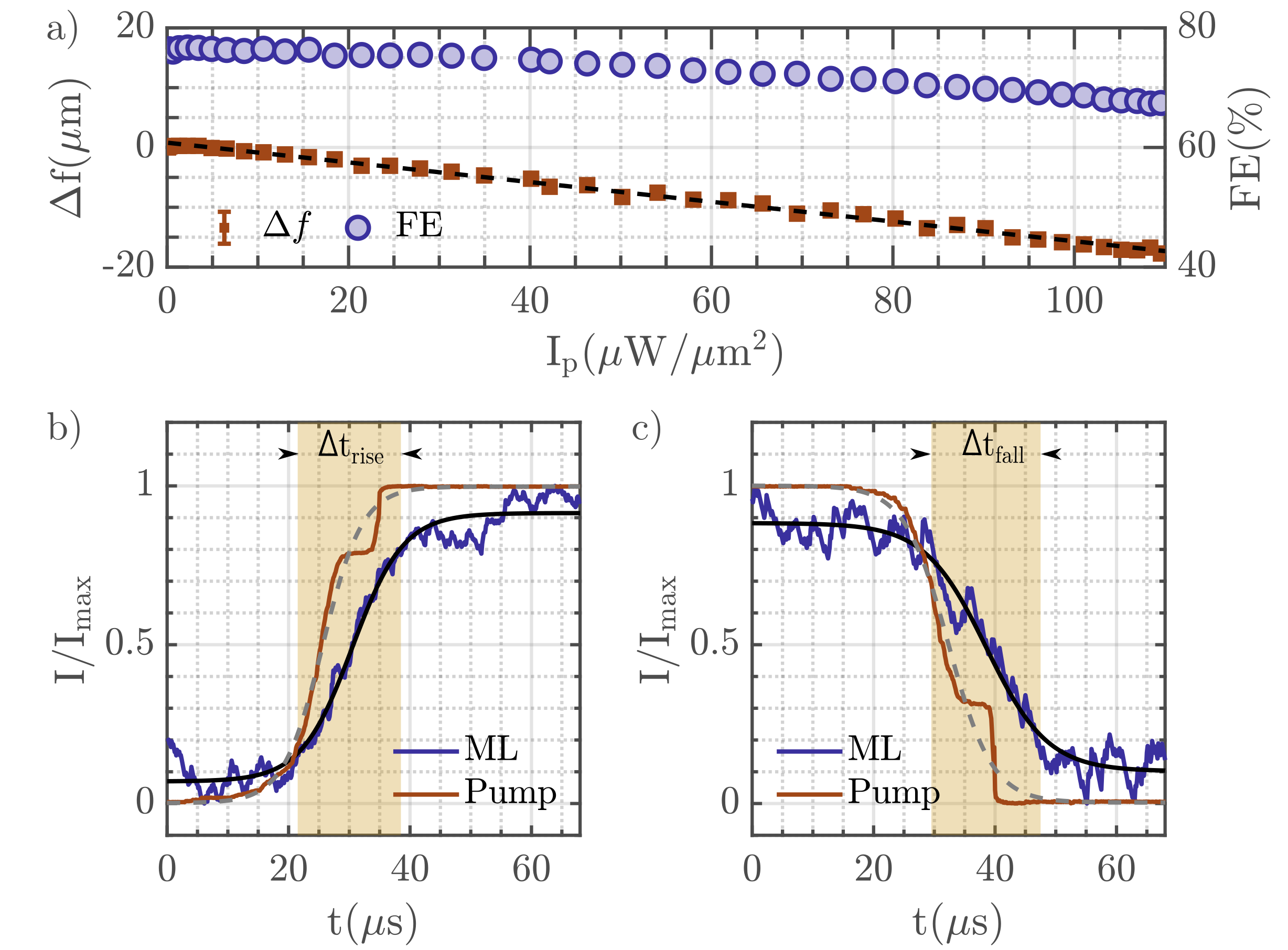}
\caption{Focal length reconfigurability and switching dynamics of the metalens. \textbf{a}) Focal length change and focusing efficiency (FE) as a function of pump intensity $I_{p}$. Red dots represent experimental data while the black solid line is a linear fit. The blue circular dots shows the focusing efficiency (FE). \textbf{b}) Time response (rise time) of the metalens (blue solid line) switching from $f$ = 320\,$\mu$m to 300\,$\mu$m under modulation of the pump beam from 0 to 110\,$\mu$W/$\mu$ m$^{2}$ (red solid line). \textbf{c}) Time response (fall time) of the metalens (blue solid line) switching from $f$= 300\,$\mu$m to 320\,$\mu$m under modulation of the pump beam from 110 to 0\,$\mu$W/$\mu$ m$^{2}$ (red solid line). In \textbf{b}) and \textbf{c}), black solid line and grey dashed lines are Sigmoid fits to the metalens and pump response, respectively. The 90\% to 10\% rise/fall time for the metalens is higlighted by the yellow shaded area.}  
\label{fig:fig5}
\end{figure*}

When sweeping the pump intensity $I_{p}$ from 0 to 110\,$\mu$W/$\mu$m$^{2}$, we observed a linear power dependence of the focal distance (Figure \ref{fig:fig5}a). Another important parameter that characterizes the performance of a metalens is the focusing efficiency (FE), defined as the relative amount of light passing through an aperture of 3 times the full width half maximum (FWHM) of the focal spot \cite{arbabi2015subwavelength}. We experimentally measured the FE for increasing $I_{p}$ and observe a reduction from 78\% (for $I_{p} = 0\,\mu$W/$\mu$m$^{2}$) to 68\% (for $I_{p} = 110\,\mu$W/$\mu$m$^{2}$) as depicted in Figure \ref{fig:fig5}a (blue circular dots).

Finally, we are interested in the switching dynamics of our optomechanically reconfigurable metalens. To measure its time response, we modulated the pump beam through the EOM with a square wave of frequency 0.5 Hz and amplitude 110 $\mu$W/$\mu$m$^{2}$. The switching on rise time (10-90\%) and switching off fall time (90-10\%) of the EOM is limited to 14 $\mu$s. Figure \ref{fig:fig5}b and c shows the time response of both the pump (red solid curve) and the metalens (blue solid line). We fit a Sigmoid function to the data (black solid line for the metalens and grey dashed line for the pump) to obtain the rise and fall time. Our measurements give a rise time $\Delta t_{rise} = 17\,\mu$s (10\% to 90\%) and fall time $\Delta t_{fall} = 18\,\mu$s (90\% to 10\%) for the metalens highlighted by yellow shaded area in Figure \ref{fig:fig5}b and c, respectively. From these values, we rule out that thermal effects dominate in the tunability of our lens. This conclusion is corroborated by the switching time calculated by Zhang et al. \cite{Zhang2013}.

In conclusion, we proposed and realized an ultra-thin varifocal metalens actuated by optomechanical control, which combines significant tunability with 10\,$\mu$s response time. While the system was optimized to operate in the NIR region of the spectrum, extension to the visible frequency range could be achieved through an appropriate adjustment of the meta-atoms parameters and choice of their constitutive material. We foresee that future development of optomechanical integrated reconfigurable metalenses and other metasurface-based functionalities would greatly benefit from further mode engineering, directly leveraging on the latest advances in dielectric metasurfaces. 
\section*{Acknowledgments}
Adeel Afridi acknowledges financial support from the Marie Skłodowska-Curie Co-funding Programme (COFUND-DP, H2020-MSCA-COFUND-2014, GA No. 665884).
\bibliography{references}
\end{document}


\maketitle
\section*{Numerical simulation}
Optical response and mechanical actuation of the meta-atoms were numerically simulated using FEM based solver (COMSOL MULTIPHYSICS). For optical analysis and optimization, the RF module of the simulator was used, while the Structural Mechanics module was used for the mechanical analysis. \\
For optical simulation, a single unit cell (see Figure 2a in the main text) consisting of the suspended silicon meta-atom ($n$ = 3.2+0.04i \cite{Karvounis2015a}) in vacuum ($\epsilon_0$ = 1) was defined. Periodic boundary conditions in $x$ and $y$ directions were used and perfectly matching layer (PML) were used in $\pm z$ direction. The unit cell was illuminated using a plane wave (port boundary condition) with normal incidence. The unit cell was meshed with a step size of $\lambda/(20n)$, where $\lambda$ is the wavelength of excitation and $n$ is the refractive index. Finally the spectral and phase response were extracted from the scattering parameters (S-parameters) of the port boundary condition.\\
\section*{Optical Force Calculation}
In classical electrodynamics, the components of the time averaged, total force \textit{F}, encompassing radiation pressure and gradient force, exerted by an optical field on a meta-atom can be calculated as a surface integral over the time averaged Maxwell's tensors \cite{jackson1998classical}:
\[<F_{i}> = \oiint_{S}<T_{ij}>n_{j}dS\]
where \textit{S} is the bounding surface around the meta-atom and $n_j$ is the unit vector pointing out of the surface. We divided the bounding surfaces into two, i.e. around the disk and around the beam (see Figure S3a). $<T_{ij}>$ is the time averaged Maxwell's stress tensor:
\[<T_{i j}>=\frac{1}{2} \operatorname{Re}[\varepsilon \varepsilon_{0}(E_{i}E_{j}^{*}-\frac{1}{2} \delta_{i j}|E|^{2})+\]
\[\mu \mu_{0}(H_{i} H_{j}^{*}-\frac{1}{2} \delta_{i j}|H|^{2})]\]
where \textit{E} and \textit{H} are the electric and magnetic near field, respectively, extracted from the full wave optical simulation from COMSOL.
\section*{Meta-atom library}
\begin{table}[h]
\centering
\resizebox{\textwidth}{!}{\begin{tabular}{|c|c|c|c|c|c|c|c|c|c|c|c|c|}
\hline
\multicolumn{13}{|c|}{Meta-atom library list}\\
\hline
r ($\mu m$) &0.150 &0.160 &0.170 &0.180 &0.190 &0.200 &0.205 &0.210 &0.220 &0.230 &0.240 &0.250\\
w$_{2}$ ($\mu m$) &0.120 &0.115 &0.110 &0.105 &0.100 &0.095 &0.095 &0.090 &0.085 &0.080 &0.075 &0.070\\
\hline
\end{tabular}}
\caption{Meta-atom library. Disk radius $r$ and beam width $w_2$ are optimized together in order to acheive 0-2$\pi$ phase change at $\lambda_{probe}$ = 1.31\,$\mu$m and enhanced optical forces at $\lambda_{pump}$ = 1.55\,$\mu$m.}
\end{table}
\newpage
\section*{Optical setup}
\begin{figure}[h]
\centering
\includegraphics[width=0.58\textwidth]{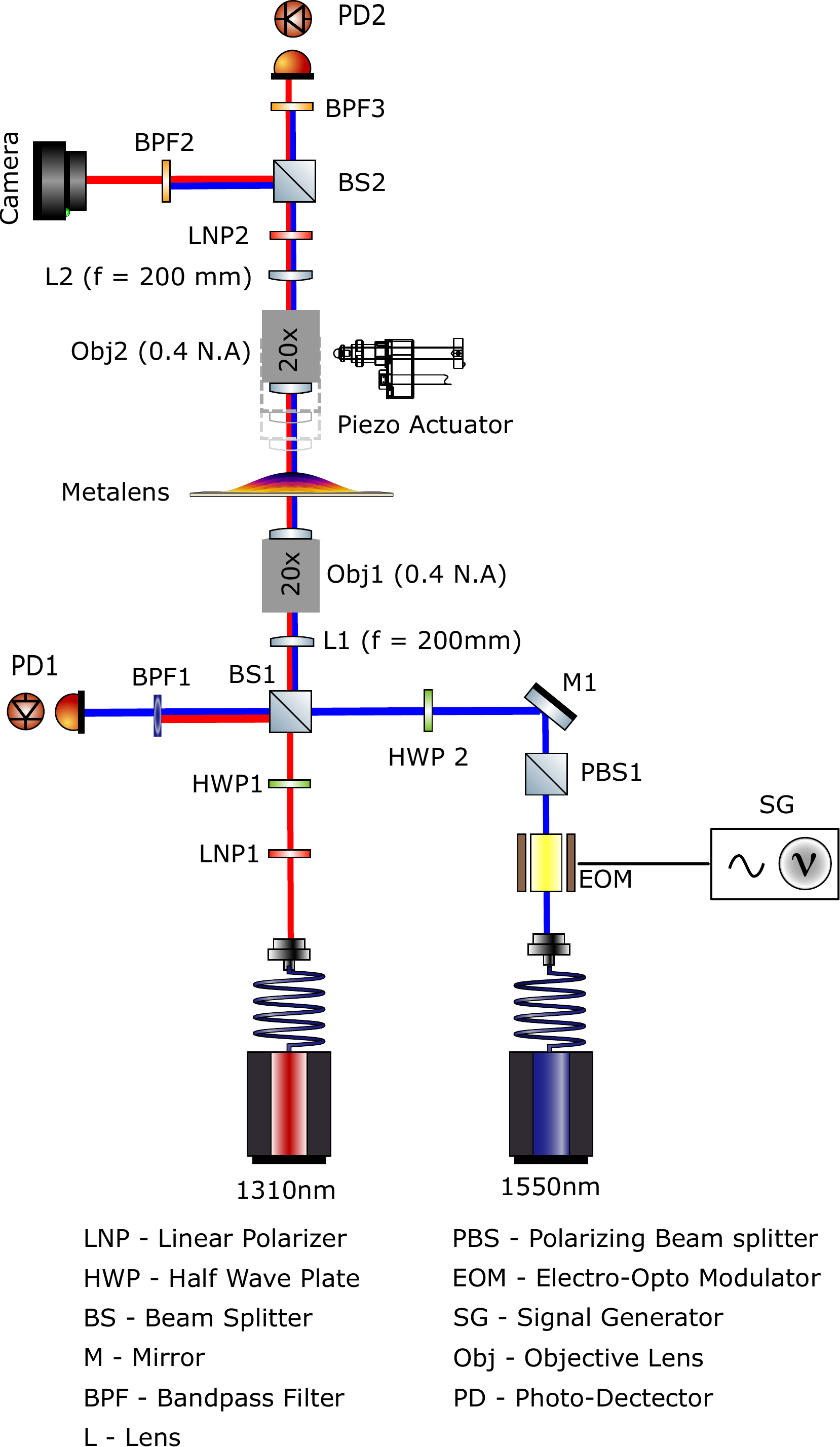}
\caption{Schematic of the two colored optical setup.}
\label{fig:S1}
\end{figure}
\newpage
\section*{Metalens performance under y-polarized pump}
\begin{figure}[ht]
\centering
\includegraphics[width=0.47\textwidth]{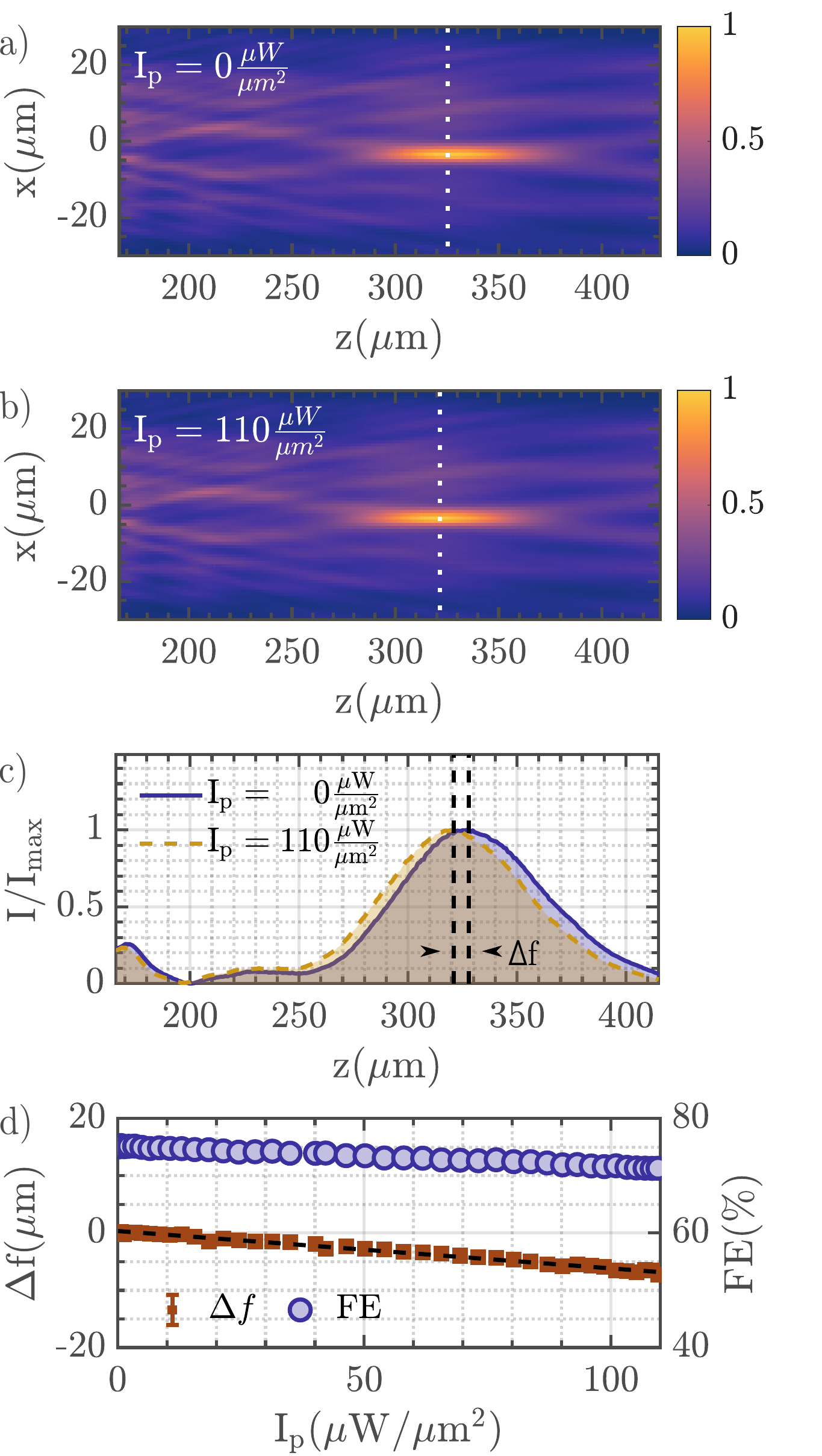}
\caption{Optical characterization of the focal length reconfigurability with pump light $y$-polarized ON and OFF. \textbf{a)} and \textbf{b)} 2D intensity maps along $xz$ propagation plane under the pump intensities $I_{p} = 0\,\mu$W/$\mu$$m^2$ and $I_{p} = 110\, \mu$W/$\mu$$m^2$, respectively. \textbf{c)} Axial optical intensity distribution for the two pump intensity states (propagation length $z$-crosscut). \textbf{d)} focal length change and focusing efficiency (FE) as a function of pump intensity $I_{p}$. Red dots represents experimental data while the black solid line is the linear fit and the blue circular dots show focusing efficiency (FE) in percentage.}
\label{fig:S2}
\end{figure}
\newpage
\section*{Optical forces and actuation under y-polarized pump}
\begin{figure}[ht]
\centering
\includegraphics[width=0.9\textwidth]{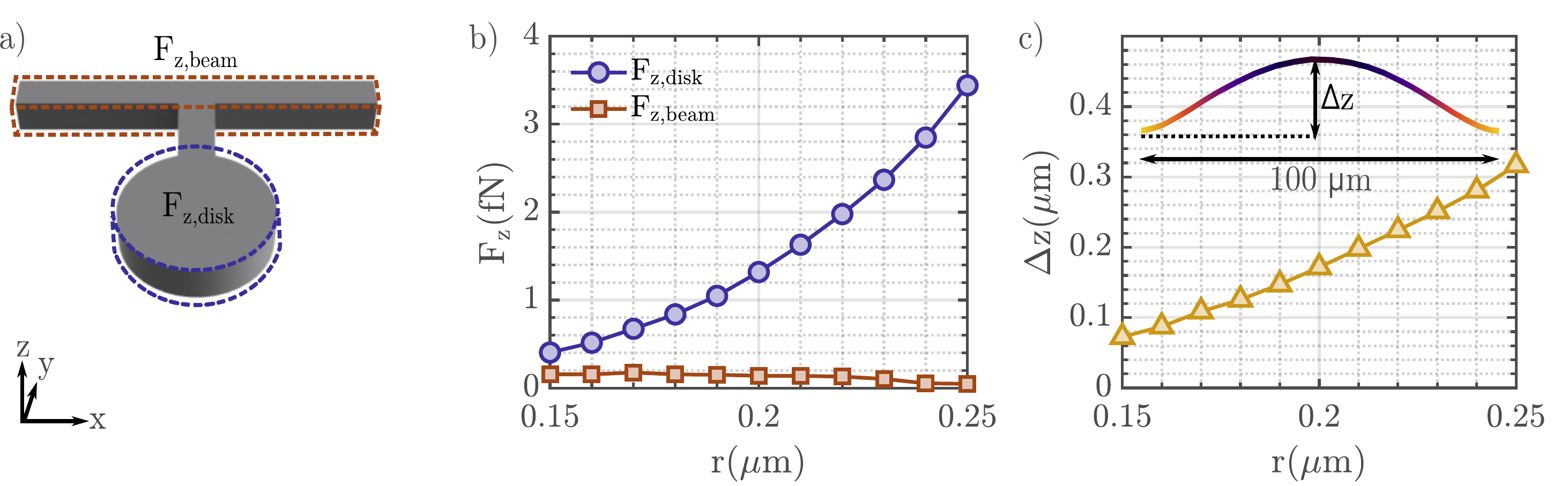}
\caption{Forces and deformation simulation results of the silicon meta-atom for $y$-polarized pump light. \textbf{a)} Meta-atom unit cell indicating bounding surfaces for the force calculation. \textbf{b)} Calculated force experienced by the meta-atoms at pump wavelength $\lambda_{probe} = 1.55\,\mu$m and pump intensity $I_{p}= 1$\,$\mu$W/$\mu$m$^2$, as a function of disk radius $r$. \textbf{c)} Numerical simulation of the deformation $\Delta z$ for a 1D meta-atom array of radius $r$ in the presence of pump light with intensity $I_p = 110\,\mu$W/$\mu$m$^2$.}
\label{fig:fig5}
\end{figure}
\bibliography{supp_references}